\documentclass[prc,aps,onecolumn,showpacs]{revtex4}
\usepackage{graphicx}
\begin{document}
\title{Low-energy $\eta$N interactions: scattering lengths
and resonance parameters}
\author{
R.~A.~Arndt, 
W.~J.~Briscoe,
T.~W.~Morrison\thanks{Present address: U.S. Department of 
                      State, Washington, DC, 20520}, 
I.~I.~Strakovsky, and
R.~L.~Workman}
\address{Center for Nuclear Studies, Physics Department \\
The George Washington University, Washington, D.C. 
20052}
\author{A. B. Gridnev}
\address{Petersburg Nuclear Physics Institute, Gatchina \\
St.~Petersburg 188350, Russia}
\draft
\date{\today}


\begin{abstract}
We consider the impact of two recent $\pi^-p\to\eta n$
measurements on the $\eta$N scattering length and $\eta$N 
branching fractions for the $N(1535)$ and $N(1520)$ 
resonances within a coupled-channel analysis of $\pi$N 
elastic scattering and $\eta$N production data. The 
sensitivity of these results to model input is also 
explored. 
\end{abstract}

\maketitle


\narrowtext
\section{Introduction}
\label{sec:intro}

While most of our knowledge of the $N$ and $\Delta$ baryons has
come from $\pi$N elastic scattering and photoproduction,
the $\eta$N channel has been crucial in determinations of the 
$N(1535)$ properties. In $\pi$N elastic scattering, this resonance 
signal is masked by a sharp cusp due to the opening $\eta$N channel, 
while in the reactions $\pi^-p\to\eta n$ and $\gamma p\to\eta p$, 
the $N(1535)$ is associated with a rapidly increasing cross section 
near threshold.  As a result, the incorporation of eta-production 
data in multi-channel fits has allowed more reliable determinations 
of the $N(1535)$ resonance parameters.

The nearby $N(1520)$ resonance, while not strongly coupled to the 
$\eta$N channel, gives an important contribution to some 
eta-production observables through interference effects.  The 
Particle Data Group~\cite{pdg} estimates the ratio $\Gamma_{\eta N} 
/ \Gamma_{\rm tot}$ to be 0.0023$\pm$0.0004, a value determined 
mainly by the multi-channel analysis of Penner and 
Mosel~\cite{penner}. In $\pi^-p\to\eta n$, the effect of this 
resonance is visible in the departure from purely S-wave behavior 
with increasing energy.  While the $N(1520)$ contribution is small, 
the effect is magnified through S-D wave interference with the 
dominant $N(1535)$ contribution. In $\eta N$ photoproduction, this 
interference effect is particularly evident in measurements with a 
polarized beam ($\Sigma$)~\cite{tiator}. 

The $\eta$N interaction has also been studied extensively due to
a strong attraction at low energies, observed in most analyses, 
which could possibly lead to the existence of bound state 
$\eta$-mesic nuclei \cite{bh85,sokol}.  Unfortunately, the $\eta N$ 
scattering length cannot be measured directly. Instead, cross 
sections for $\eta$ meson production, $\pi^-p\to\eta n$ and $\gamma 
p\to\eta p$, have been studied
~\cite{e909,pr05,bi73,bu69,de75,de69,ri70,kr95,pr95,re02,dl69,dy95,er68,du02,cr05}. 
In Ref.~\cite{bi96}, it was demonstrated that the real part of the 
scattering length cannot be extracted directly from the low-energy  
eta-production cross sections and a model analysis is needed.  
Previous analyses
~\cite{bi96,ka97,bi73,be91,bh85,kr94,gr00,ca00,ba95,si02,kr00,br02,fa95,ti94,fe98,sa95,wi97,kr01,wi93,ab96,ka95,ba97,ba98,gr97,ra01,ni01,tu65,gr99,green,ar92,penner,fi02,ba95a}
have found a large spread for the real part, from a negative value
\cite{bi96} to 1~fm \cite{ar92} (Table~\ref{tbl1}).  Reasons for 
the spread include the rather old and conflicting $\eta$ meson 
production data and differing $\pi N$ elastic scattering amplitudes 
from the Karlsruhe~\cite{kh80} and SAID~\cite{sm95} groups. However, 
the largest factor appears to be the model used in analyzing the 
data. 

In recent years, the eta-production database has seen significant 
improvements.  In this paper, we present the results from analyses 
of these new data. The most recent data for the reaction $\pi^-p\to 
\eta n$ are described in Section~\ref{sec:exp}. In 
Section~\ref{sec:ca}, we described the formalism associated with 
our fits. Results and comparisons with previous determinations are 
tabulated in Section~\ref{sec:andi}.  Finally, in 
Section~\ref{sec:conc}, we summarize our findings and consider the 
open questions which will require further work.

\section{The Experimental Database}
\label{sec:exp}

Until recently, the $\pi^-p\to\eta n$ cross section database 
contained mainly old and often conflicting measurements (see 
Ref.~\cite{fa02} for details), with no polarized measurements 
existing below 1.1~GeV/c~\cite{isaid}.  These older data have been 
reviewed by Clajus and Nefkens~\cite{clnef}. More recently, 
differential and total cross sections for $\pi^-p\to\eta n$ near 
threshold have been measured using the BNL-AGS Facility 
(E909)~\cite{e909}.  Data were obtained from threshold ($p_{\pi}$ = 
684.5~MeV/$c$) to $\sim$770~MeV/$c$.

\begin{figure*}[th]
\centering{
\includegraphics[height=0.45\textwidth, angle=90]{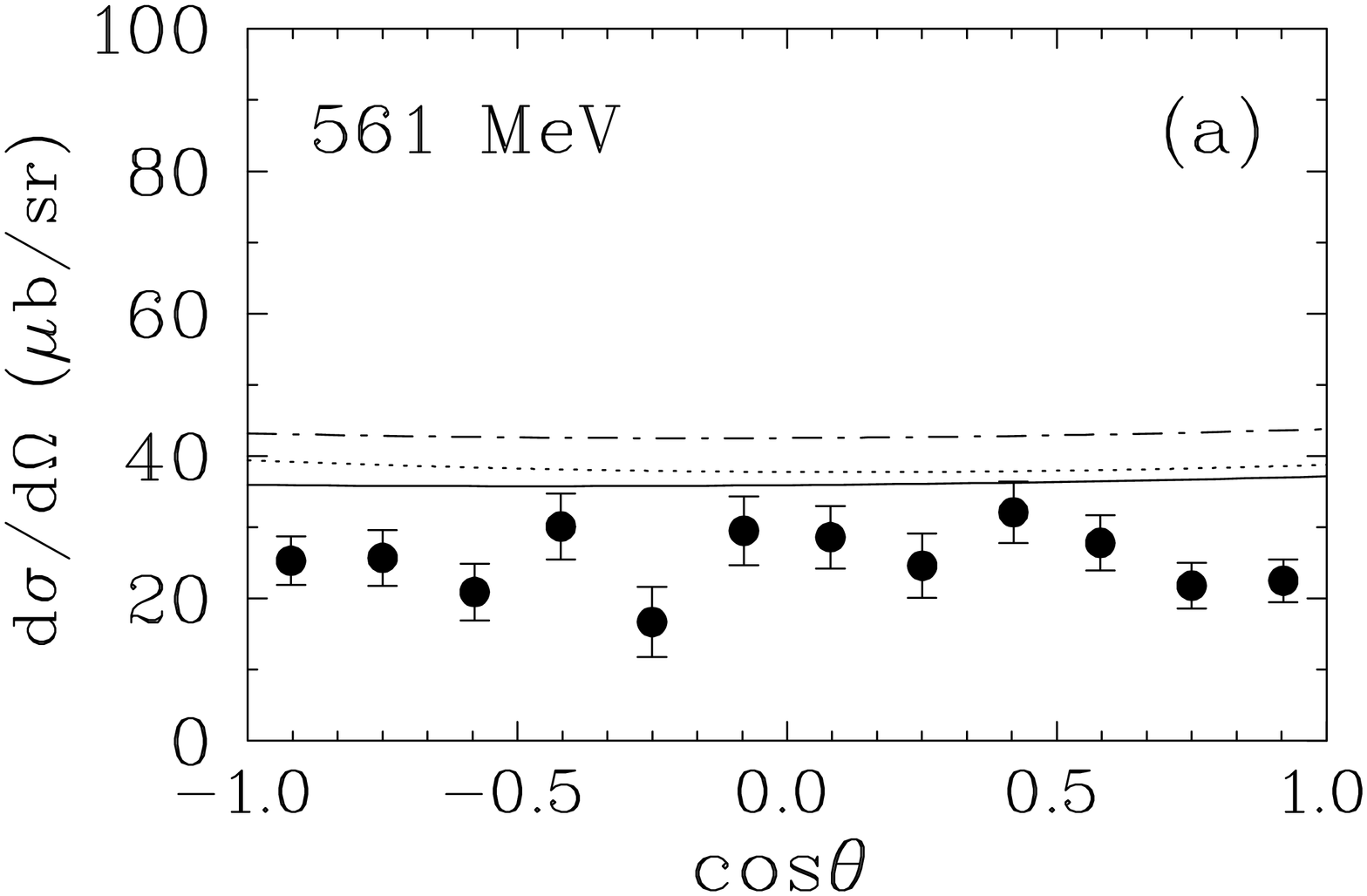}\hfill
\includegraphics[height=0.45\textwidth, angle=90]{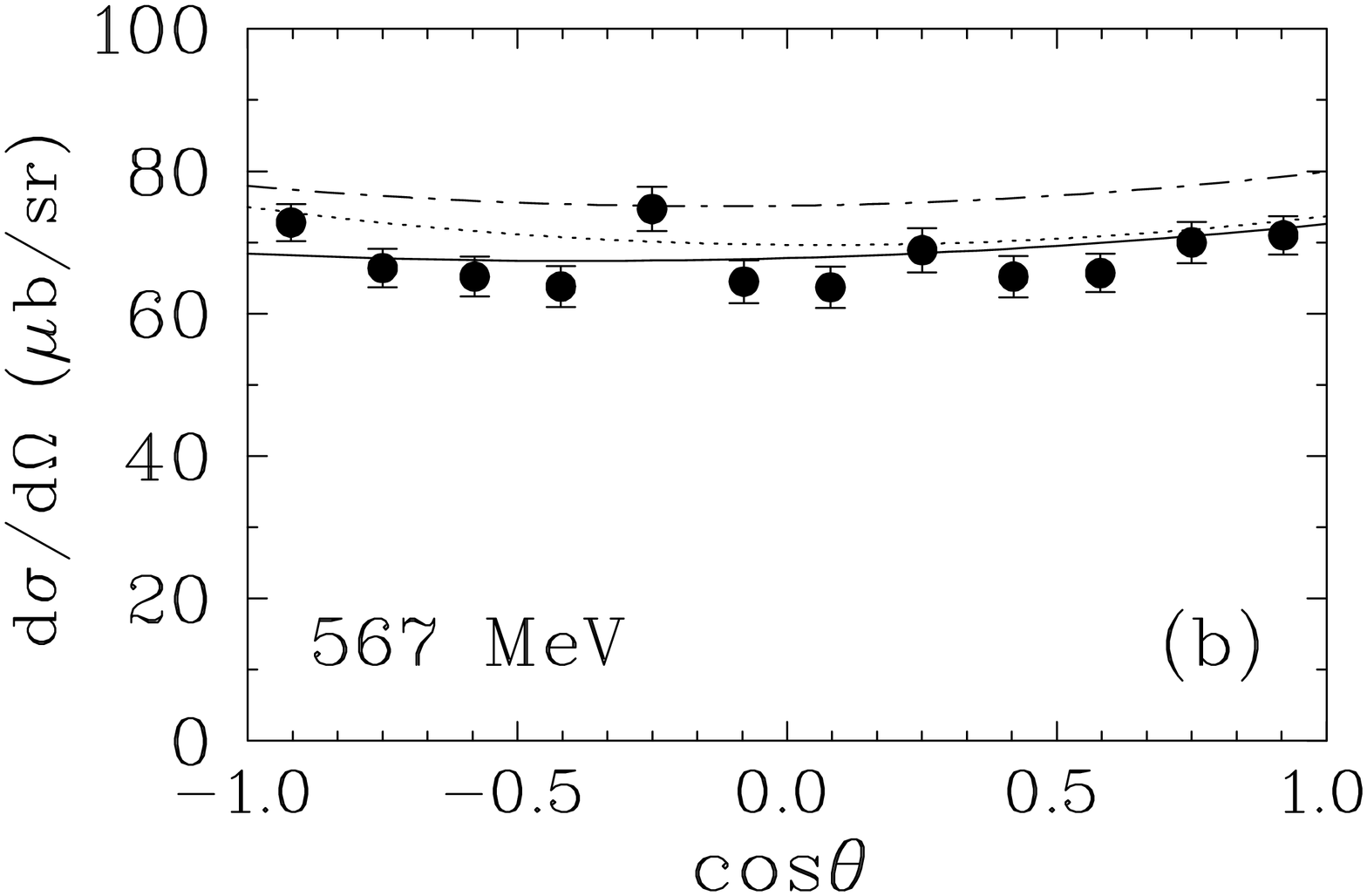}
\includegraphics[height=0.45\textwidth, angle=90]{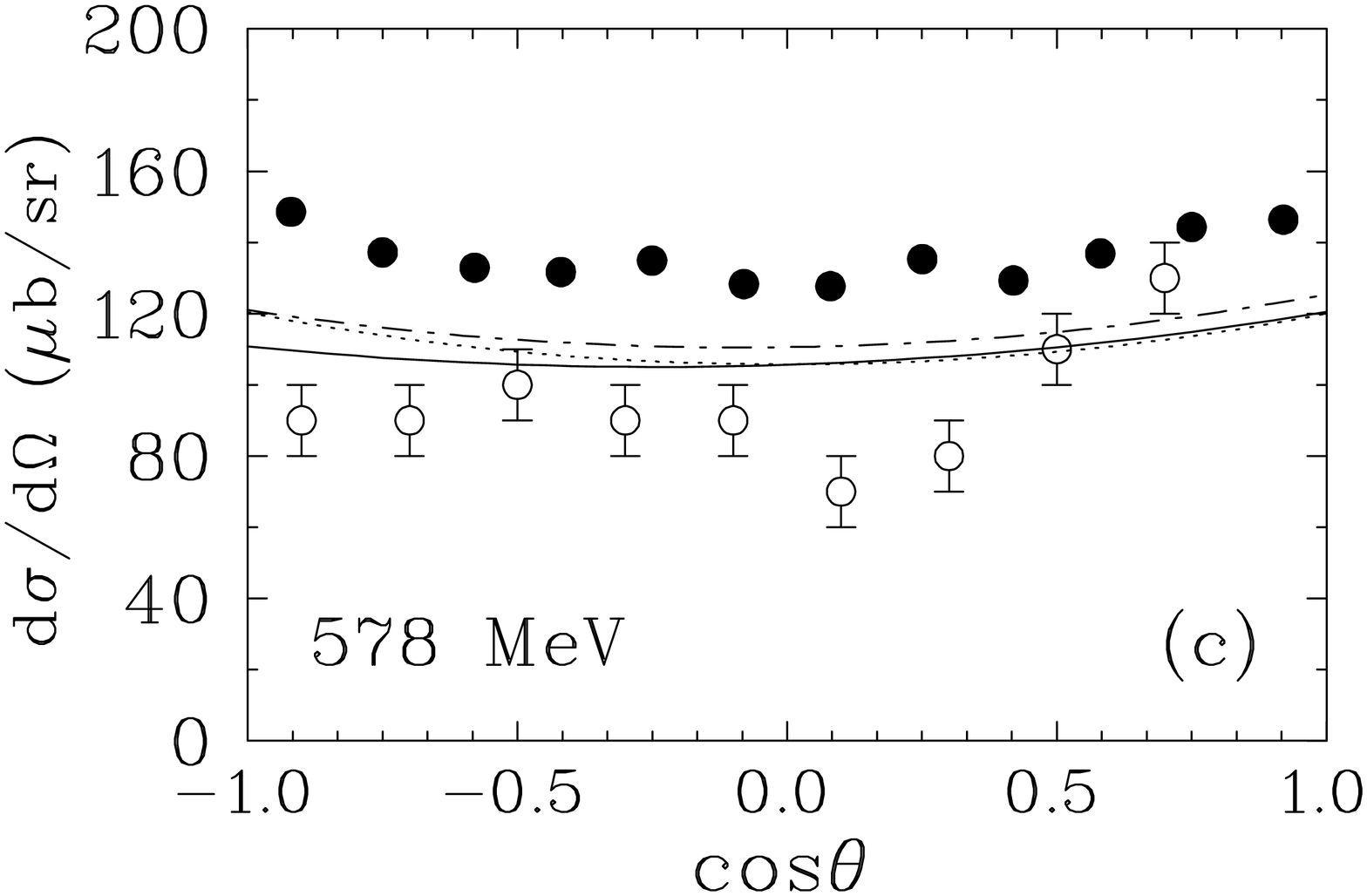}\hfill
\includegraphics[height=0.45\textwidth, angle=90]{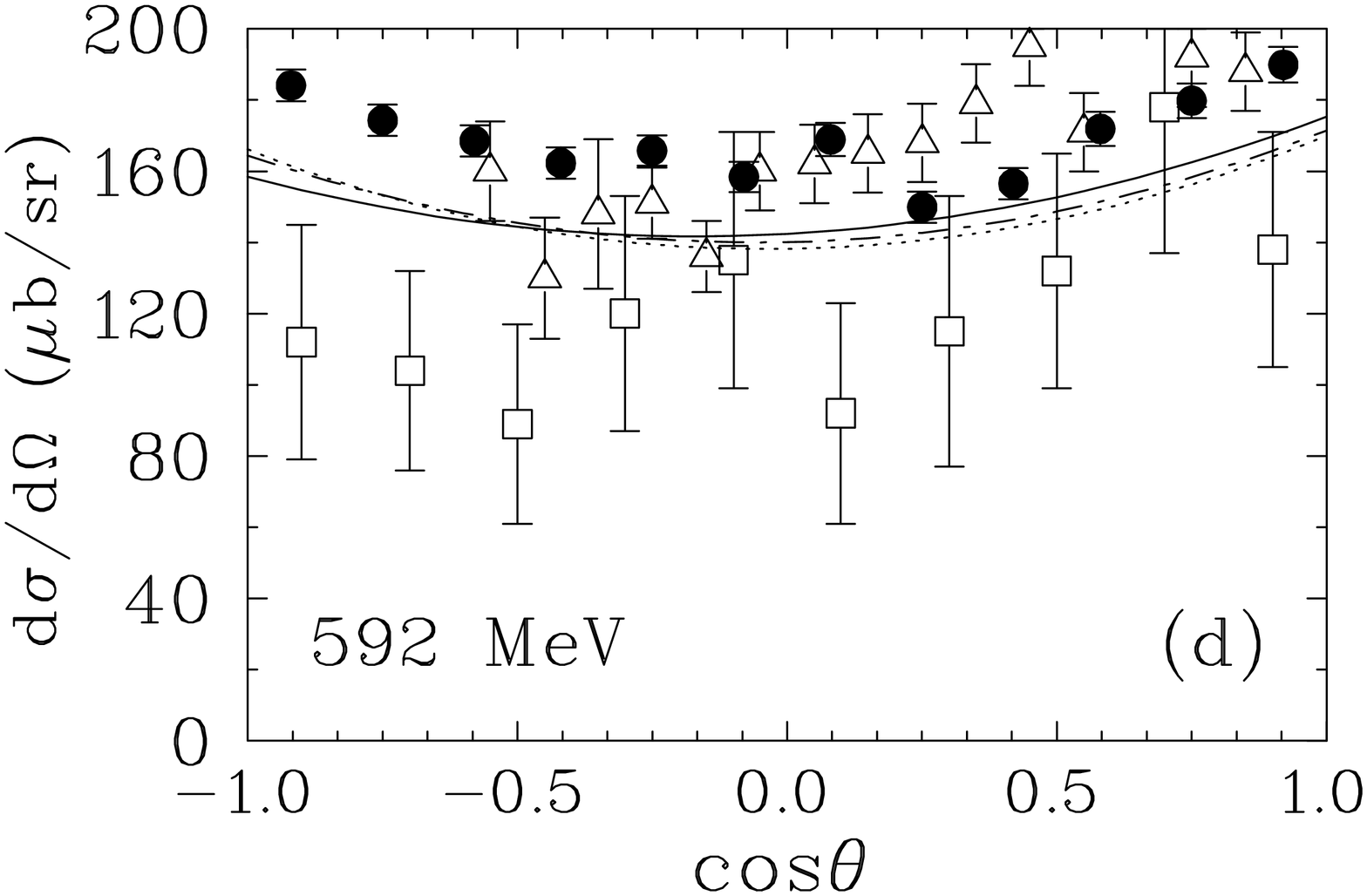}
\includegraphics[height=0.45\textwidth, angle=90]{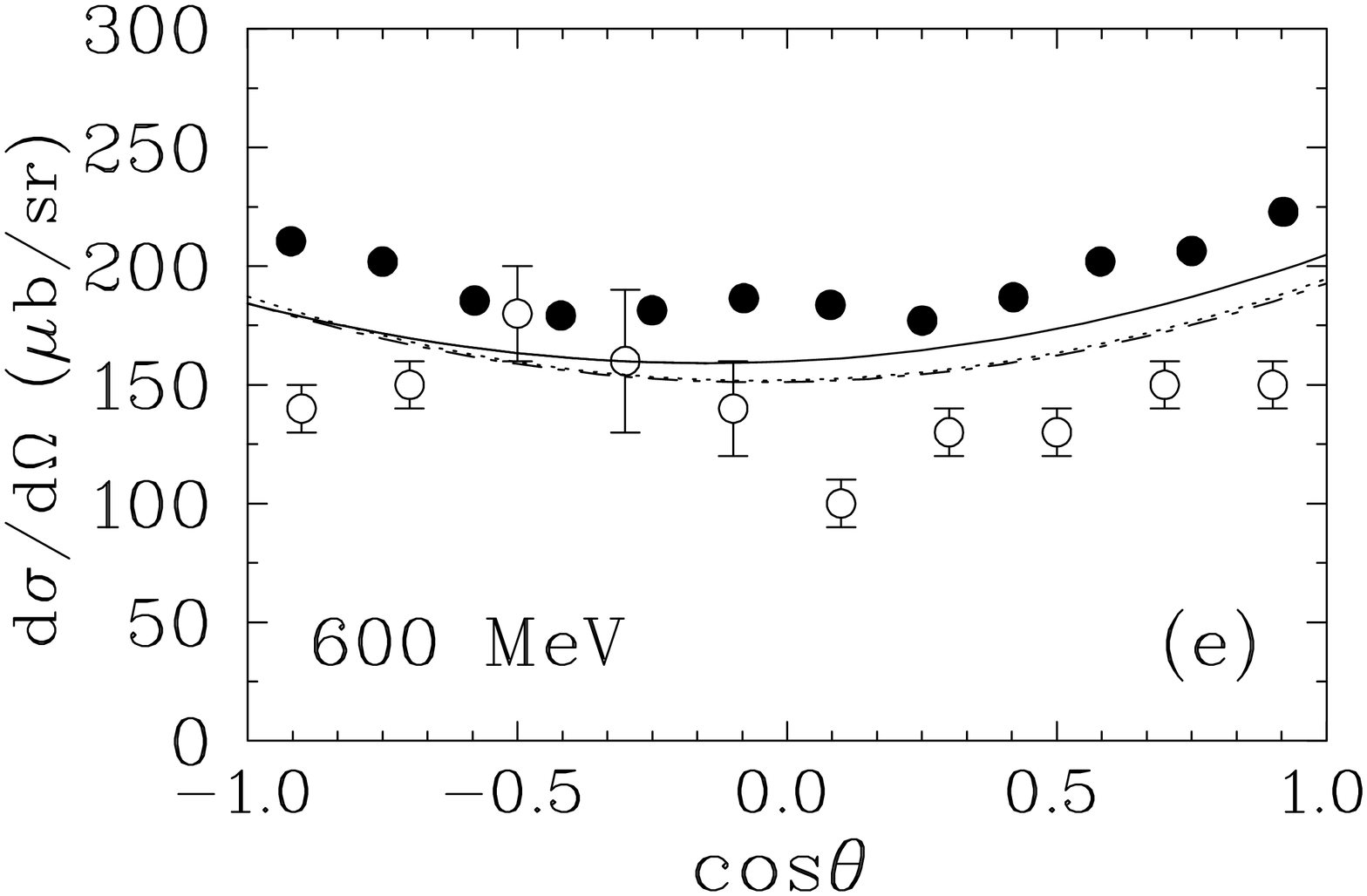}\hfill
\includegraphics[height=0.45\textwidth, angle=90]{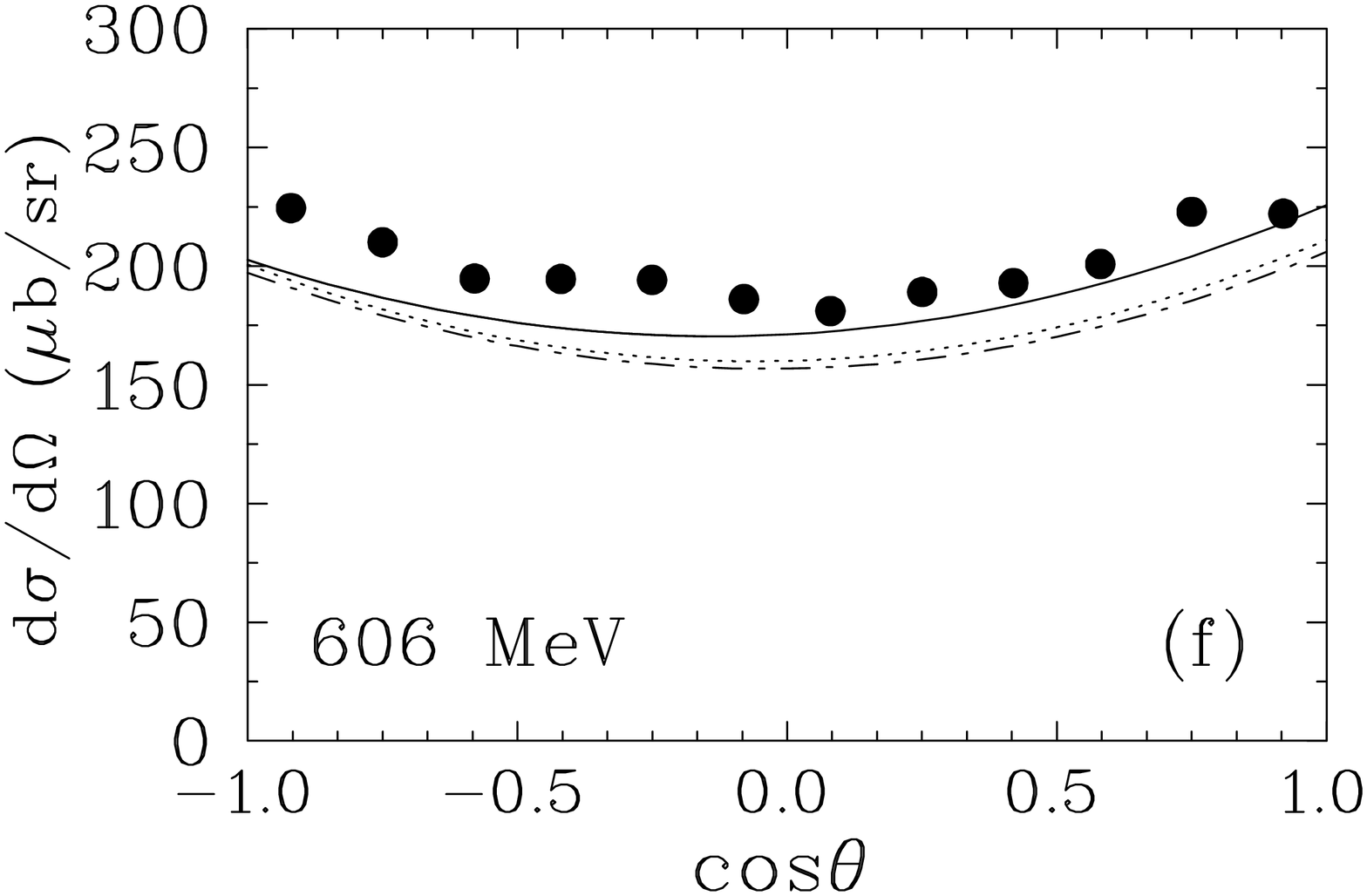}
\includegraphics[height=0.45\textwidth, angle=90]{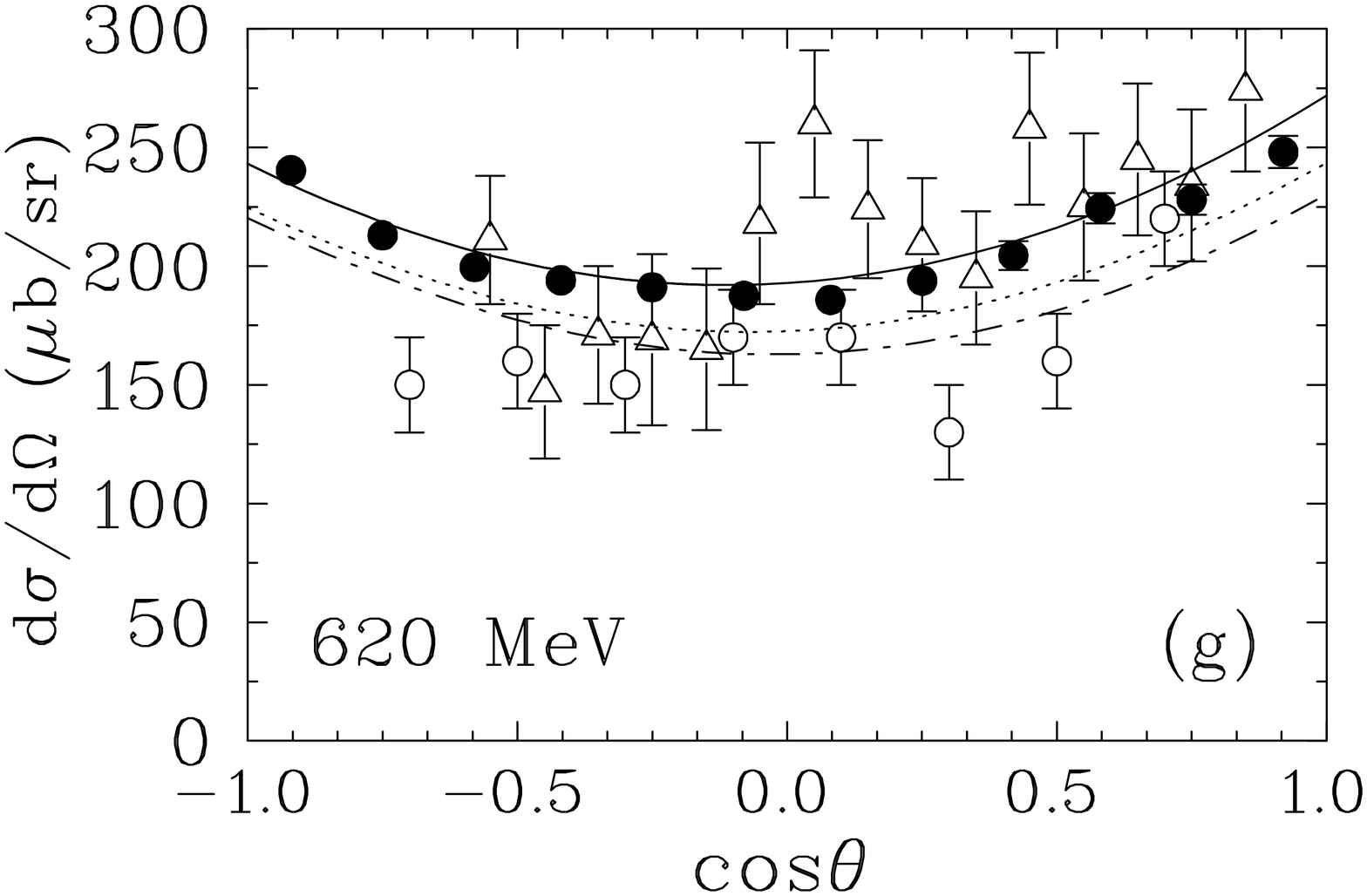}\hfill
\includegraphics[height=0.45\textwidth, angle=90]{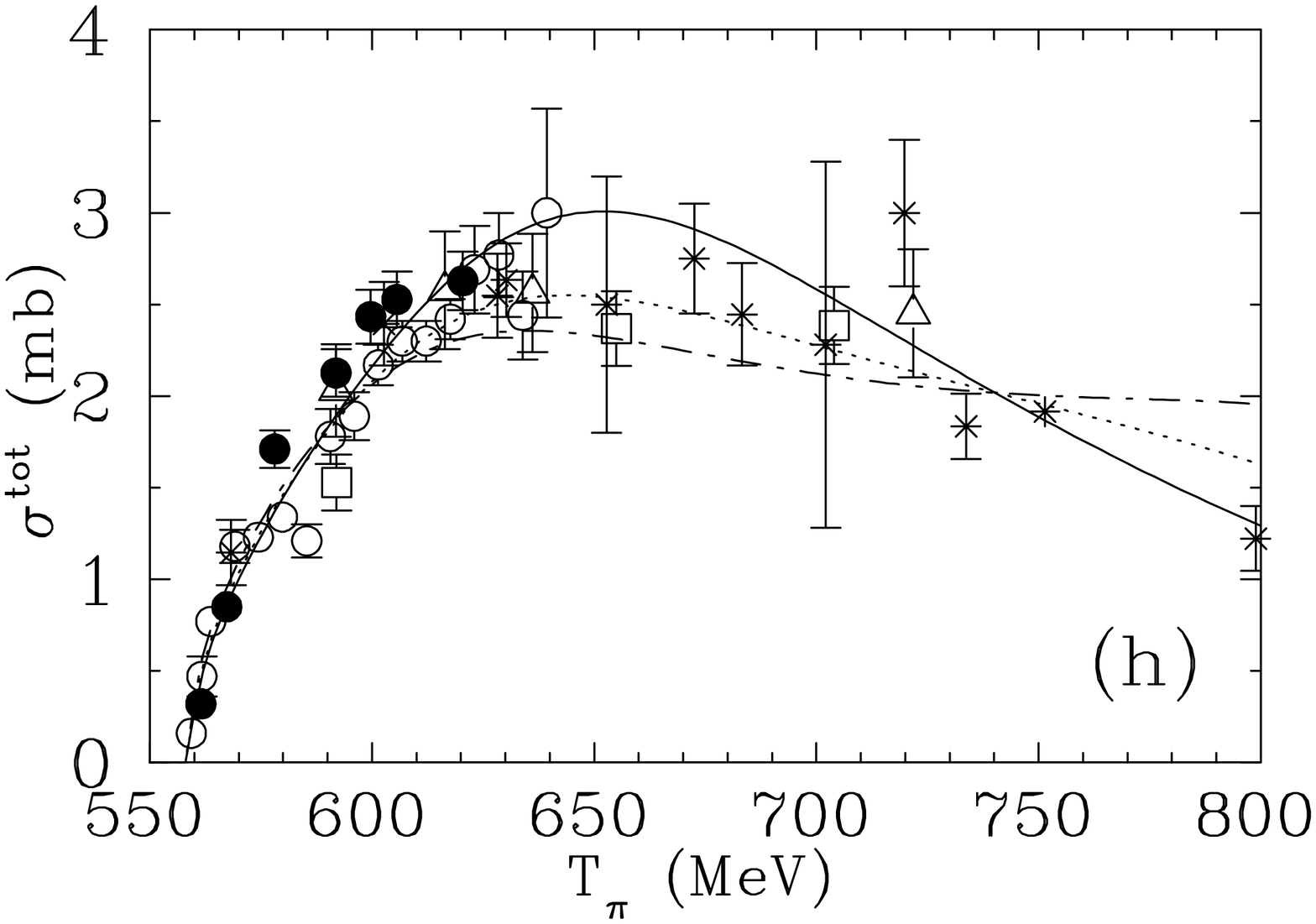}
}\caption{(a)--(g) Differential cross sections for
          $\pi^-p\to\eta n$ at seven incident $\pi^-$
          energies.  The uncertainties are statistical only.
          For the total cross sections (h), we have combined
          statistical and systematic uncertainties in
          quadrature.  
          FA02~\protect\cite{fa02} (E913/E914 data not included),
          G380, and Fit~A  shown as 
          solid, dash-dotted, and dotted lines, respectively.
          Experimental data are from 
          ~\protect\cite{pr05} (filled circles),
          ~\protect\cite{e909} (open circles),          
          ~\protect\cite{de69} (open triangles), and
          ~\protect\cite{ri70} (open squares) measurements.
          Other previous measurements (for references see 
          SAID database~\protect\cite{isaid}) shown as 
          asterisks.
} \label{fig:a1}
\end{figure*}

The most recent $\pi^-p\to\eta n$ experiment was also performed 
at BNL, but in this case, using the Crystal Ball spectrometer 
(now moved to MAMI at Mainz).  Cross sections were measured 
from threshold to 747~MeV/$c$ (E913/914)~\cite{pr05}.  The total 
and differential cross sections from these two BNL measurements 
are compared in Fig.~\ref{fig:a1}.  Normalization issues still 
remain for the differential cross sections. However, the 
Crystal Ball distributions show a much smoother variation and 
a clear onset of higher partial-wave contributions with 
increasing energy.  This feature, which is a vast improvement 
over previous measurements, allows an improved separation of 
the $N(1520)$ contribution.

Traditionally, the total cross-section is plotted as a function 
of the pion laboratory energy [Fig~\ref{fig:a1}(h)].  This view 
shows the sharp growth above threshold which is usually attributed 
to the dominance of the $N(1535)$ resonance, having a mass close 
to the $\eta$ production threshold ($\sqrt{s}$ = 1487~MeV) and a 
strong coupling to the $\eta N$ system.  In Fig.~\ref{fig:a2}, we 
instead plot the total cross-section as a function of the $\eta$ 
cm momentum $p^\ast_{\eta}$.  From the figure, we see that data 
can be described very well by a linear fit (dashed line).  This 
is due to the $S$-wave dominance of the total cross section.  
From the slope of the best-fit line, a restriction on the 
imaginary part of the $\eta N$ elastic scattering amplitude 
$A_{\eta N}$ can be found.  

The optical theorem leads to
\begin{eqnarray}
ImA_{\eta N}=\frac{p_{\eta}^\ast}{4 \pi}~\sigma_{\eta n}^{tot}
\phantom{xxxxxxxxx}\nonumber \\ 
= \frac{p_{\eta}^\ast}{4\pi}~(\sigma_{\eta n\to\pi N} + 
\sigma_{\eta n\to 2\pi N} + \sigma_{\eta n\to \eta n})
\phantom{xxxx}\nonumber \\
= \frac{3~p_{\pi}^{*2}}{8 \pi p_{\eta}^\ast}
~\sigma_{\pi^-p\to\eta n} + \frac{p_{\eta}^\ast}{4 
\pi}~(\sigma_{\eta n\to 2\pi N} + \sigma_{\eta n\to 
\eta n}).
\label{2} \end{eqnarray}
As a result, we have
\begin{eqnarray}
ImA_{\eta N}\ge \frac{3 {p_{\pi}^\ast}^2}{8 \pi 
p_{\eta}^\ast}~\sigma_{\pi ^- p\to\eta n}.
\label{3} \end{eqnarray}

Using a linear fit, the recent E909 threshold data~\cite{e909} 
give
\begin{eqnarray}
\frac{1}{p_{\eta}^\ast}\sigma_{\pi ^- p\to\eta n} 
= 15.2 \pm 0.8~{\rm\mu b/MeV}
\nonumber \\
ImA_{\eta N}\ge 0.172 \pm 0.009~{\rm fm},
\phantom{xxxx}\label{4} \end{eqnarray}
which can be compared with a previous output from 
Ref.~\cite{bi73}
\begin{eqnarray}
\frac{1}{p_{\eta}^\ast}\sigma_{\pi^-p\to\eta n}
= 21.2 \pm 1.8 ~{\rm\mu b/MeV}~{\rm and}~
\nonumber \\
ImA_{\eta N}\ge 0.24 \pm 0.02~{\rm fm}.
\phantom{xxxxx}\label{5} \end{eqnarray}

It is commonly believed that the $N(1535)$ resonance 
dominates the $\eta$ production cross section.   This 
resonance mechanism results in the imaginary part of the 
$\eta$ pion-production and the $\eta N$ elastic scattering 
amplitudes being determined mainly by the $N(1535)$ resonance 
parameters.  But this is not the case for the real part.  The 
real part of the resonance amplitude goes to zero at the 
resonance position.  Therefore, the real part of the $\eta N$ 
scattering length strongly depends on nonresonant processes. 
For this reason, a multichannel analysis is favored in 
determining the $\eta N$ scattering length. Our approach is 
described in the next section.

\section{Combined analysis of $\pi N$ elastic and $\pi^-p\to\eta 
n$ data}
\label{sec:ca}

Our energy-dependent partial-wave fits are parametrized in 
terms of a coupled-channel Chew-Mandelstam K matrix, as 
described in Ref.~\cite{fa02}.  This choice determines the way 
we modify energy dependence and account for unitarity in our 
fits. Data for $\pi N$ elastic scattering have been fitted up to 
2.1~GeV in the pion lab kinetic energy. Data for the reaction 
$\pi^-p\to\eta n$ have been included from threshold up to 0.8~GeV.   
Constraint data have also been included, in order to ensure that 
the resulting fit produces elastic $\pi N$ amplitudes satisfying 
a set of forward and fixed-t dispersion-relations.  This fit to 
data plus constraints must be iterated until a stable result is 
obtained~\cite{fa02}.  Finally, we have included $\pi \Delta$ 
and $\rho N$ channels to account for unitarity, but have not 
explicitly fitted data of this type. 

The $N(1535)$ resonance couples mainly to $\pi N$ and $\eta N$, 
with a much smaller branching fraction to $\pi\pi N$, and our 
results were not sensitive to the choice of additional channels. 
For the $N(1520)$, however, there is a substantial inelastic 
branching to $\pi\pi N$, split mainly between $\rho N$ and 
$\pi\Delta$.  We therefore considered two different fits having 
(a) approximately equal $\rho N$ and $\pi\Delta$ branching 
fractions, and (b) a larger $\rho N$ branching fraction. While 
this choice had little effect on the total width, it significantly 
changed the branching to $\eta N$.

In order to extract resonance parameters from our global fits, we have
generally extrapolated into the complex energy plane to search for 
poles.  We have also fitted our energy-dependent and single-energy 
partial-wave amplitudes with Breit-Wigner plus background forms. 
Here we have chosen to fit the partial waves containing the 
$N(1535)$ and $N(1520)$ resonances in terms of a K-matrix 
resonance form, allowing for 2 poles in the $S_{11}$ partial wave, 
plus background. This revised parametrization was then fitted
directly to the data (from 400 to 900~MeV in the pion kinetic 
energy) in order to determine resonance parameters.  The remaining 
partial waves were fixed to values determined from a previous global 
(energy-dependent) fit. This resulted in error estimates more 
directly tied to the data. 

The general form used for the modified ($S_{11}$ and $D_{13}$) 
partial waves was 
\begin{equation}
T \; = \; \rho^{1/2}\; T_x \; \rho^{1/2} ,
\end{equation}
with $\rho^i$ giving the phase space for a channel ($\pi N$, 
$\pi\Delta$, $\rho N$, or $\eta N$), and with $T_x$ represented 
in terms of a K-matrix as 
\begin{equation} 
T_x \; = \; K_x \left( 1 + i K_x \right)^{-1} ,
\end{equation}
where, in a two-resonance case, we have fitted
\begin{equation}
K_x \; = \; K_b \; + \; {{K_1}\over {W_1 -W}} \; + \; {{K_2}\over 
{W_2 - W}}.
\end{equation}
The background has been parametrized in terms of the phase 
space, $K^{ij}_b$ = $(\rho^i \rho^j)^{1/2} \kappa^{ij}$, with 
$\kappa^{ij}$ elements assumed constant over the limited energy 
ranges of these fits.  The K-matrix pole residues were similarly 
parametrized as $K_1^{ij}$ =$\gamma^i \gamma^j$ with $\gamma^i$
=$(\rho^i \Gamma^i/2 )^{1/2}$. The phase phase factors were 
normalized to unity at the resonance position $(W_1 )$.

\begin{figure*}[th]
\centering{
\includegraphics[height=0.45\textwidth, angle=90]{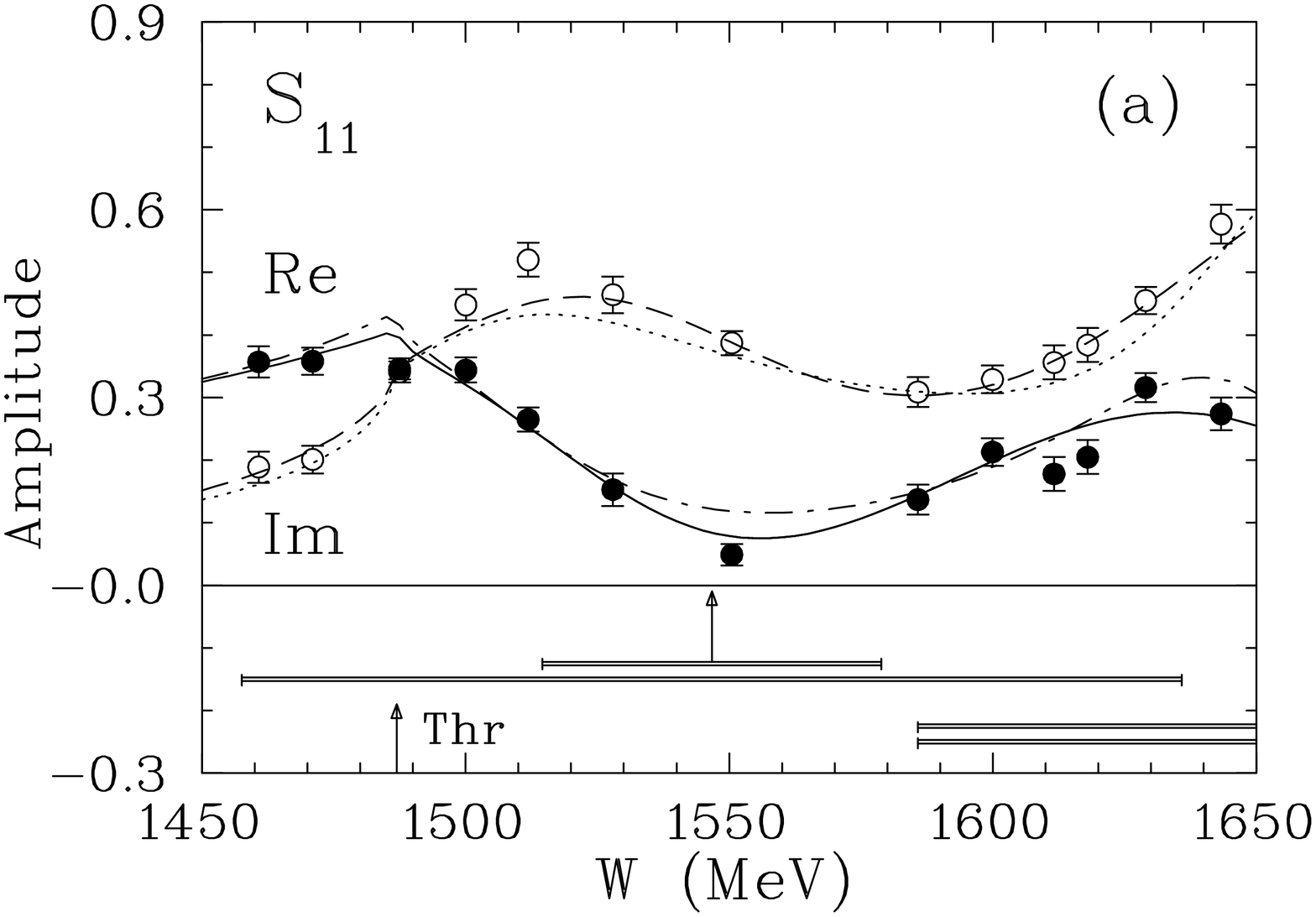}\hfill
\includegraphics[height=0.45\textwidth, angle=90]{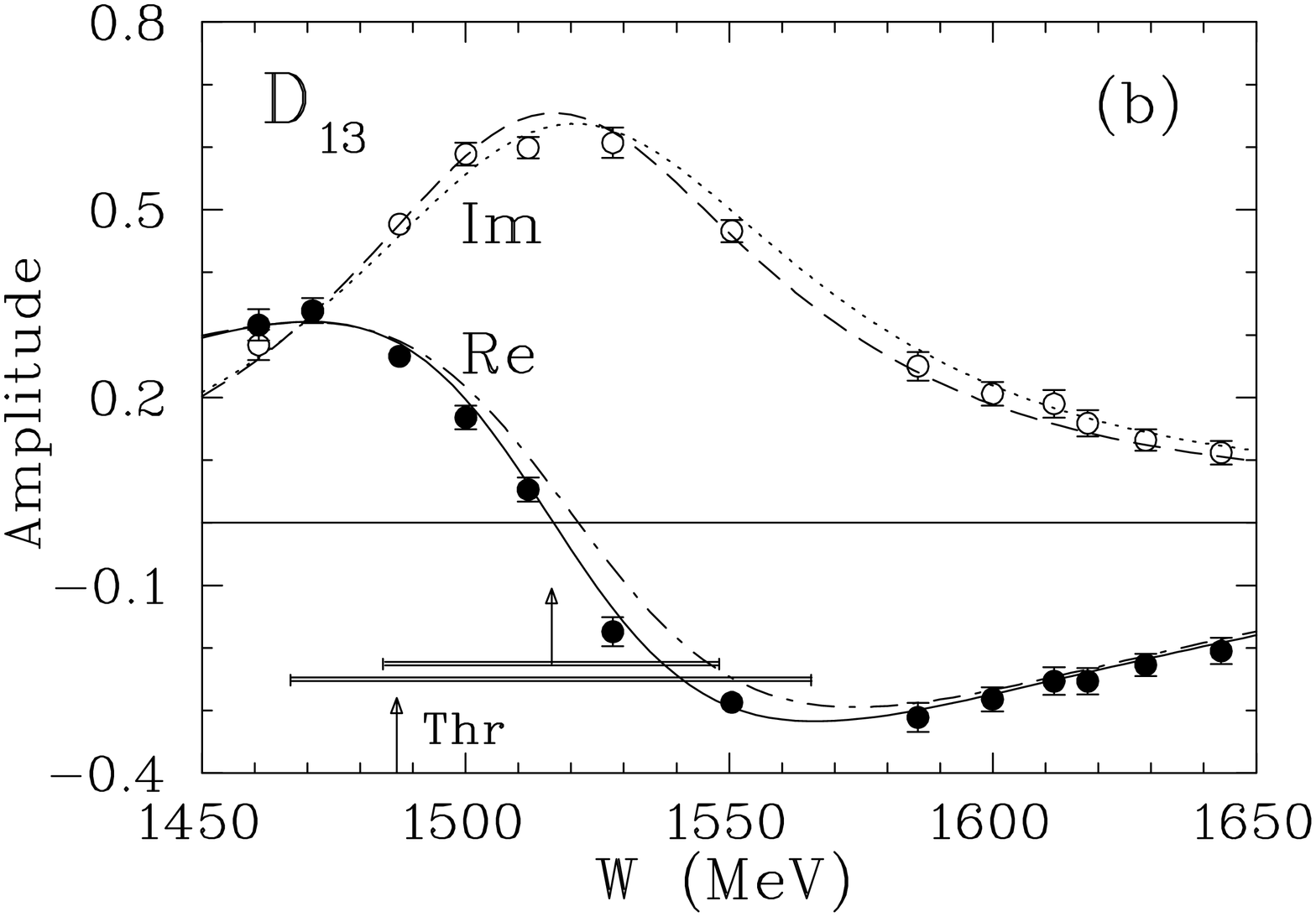}
}\caption{(a) $S_{11}$ and (b) $D_{13}$ partial amplitudes for 
          $\pi N$ elastic scattering.
          Solid (dashed) curves give the real (imaginary)           
          parts of amplitudes corresponding to the predictions of
          solution FA02~\protect\cite{fa02} (E913/E914 data not 
          included).  
          Single-energy solutions associated with FA02 are 
          plotted as filled and open circles.
          Dash-dotted (dotted) curves show the real (imaginary)         
          parts of amplitudes corresponding to G380.
          Differences between G380 and Fit~A are
          not significant.  All amplitudes are dimensionless.
          Vertical arrows indicate $W_R$ and horizontal bars show 
          full $\Gamma$/2 and partial widths for $\Gamma_{\pi N}$
          associated with the FA02 results.} \label{fig:a3}
\end{figure*}

\begin{figure*}[th]
\centering{
\includegraphics[height=0.45\textwidth, angle=90]{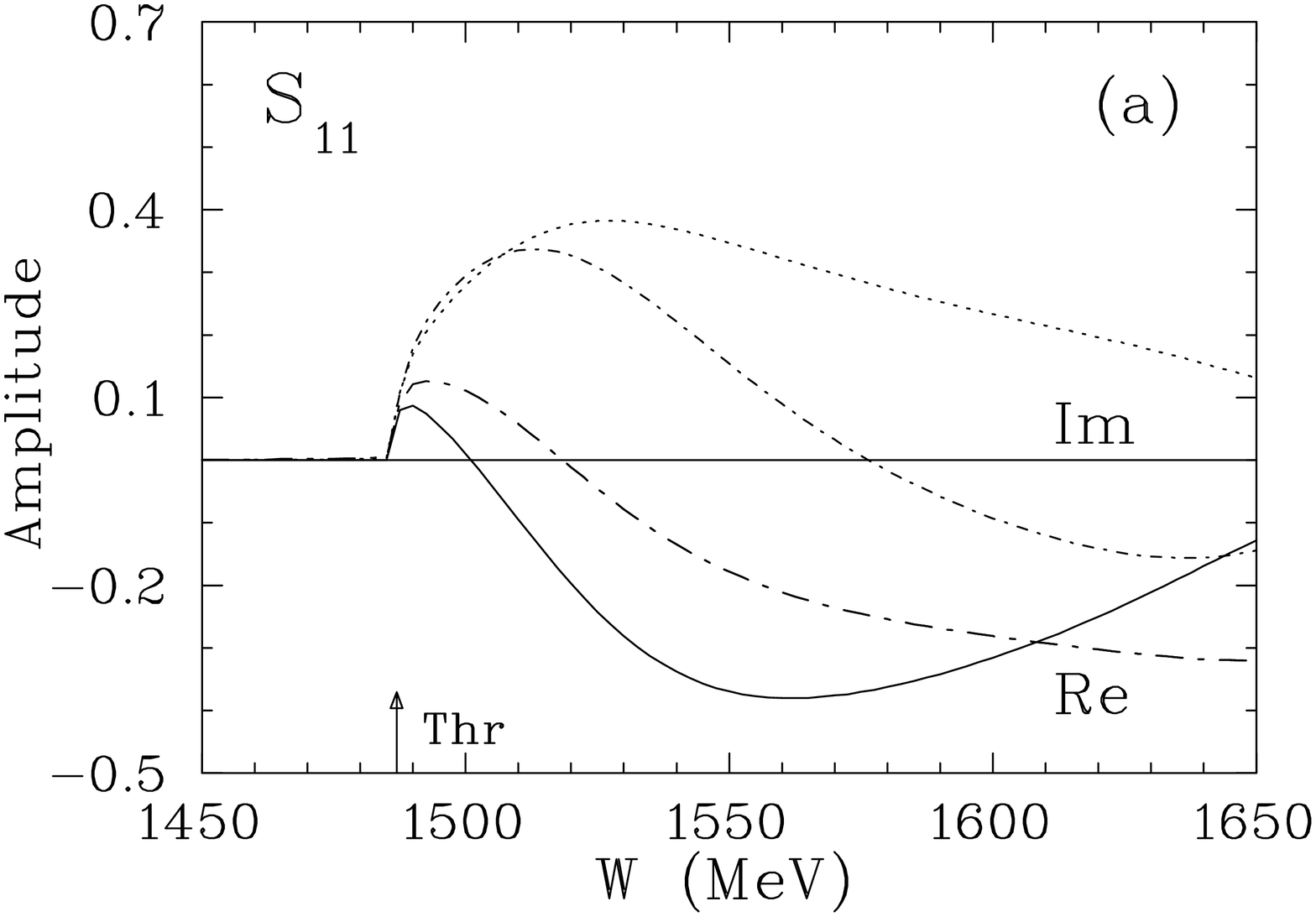}\hfill
\includegraphics[height=0.45\textwidth, angle=90]{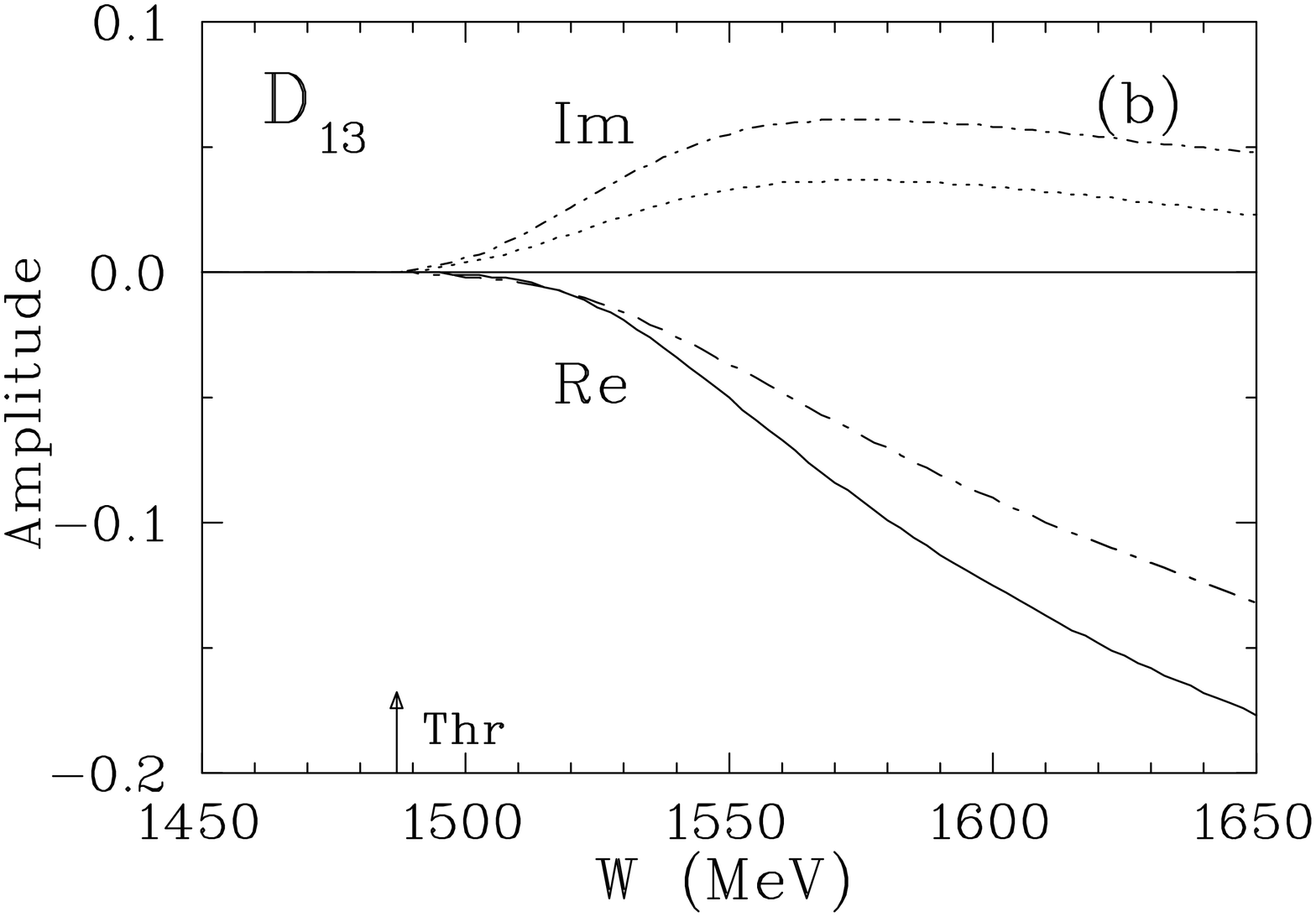}
}\caption{(a) $S_{11}$ and (b) $D_{13}$ partial amplitude for 
          $\pi^-p\to\eta n$.  Dash-dotted (dotted) curves show 
          the real (imaginary) parts of amplitudes corresponding 
          to G380.  Solid (short-dash-dotted) lines 
          represent the real (imaginary) parts of amplitudes 
          corresponding to the Fit~A.  All amplitudes are 
          dimensionless.} \label{fig:a4}
\end{figure*}

Results for the $S_{11}$ and $D_{13}$ $\pi N$ elastic
scattering amplitudes are displayed in Fig.~\ref{fig:a3}. Here 
the result of our most recently published fit (FA02) is compared 
to an updated and improved version (G380). The K-matrix fits 
closely follow the G380 result and have not been plotted. 
Fig.~\ref{fig:a4} shows the much larger deviations existing 
between different versions of the $\pi^-p\to\eta n$ amplitudes.

\section{Results and Discussion}
\label{sec:andi} 

\subsection{$\eta N$ couplings}
\label{sec:pwaa}

Our results from four fits, two with and two without the recent
Crystal Ball data, are summarized in Tables~\ref{tbl2},\ref{tbl3}. 
Listed are the partial widths for the $N(1535)$ 
and $N(1520)$ resonances.  While the PDG quotes~\cite{pdg} a 
broad and conservative range of about 30 to 50 percent for both 
the $\pi N$ and $\eta N$ branching ratios corresponding to the 
$N(1535)$, most recent determinations have found the $\eta N$ 
fraction to be about 50\%, the remaining 50\% divided between 
the $\pi N$ and $\pi\pi N$ channels. We have similarly found 
$\eta N$ branching fractions exceeding the $\pi N$ fraction in 
all of our fits.  The $N(1535)$ total width, found in the 
K-matrix fits, differs significantly from our previously 
published result~\cite{fa02}. This is due to the coupled-channel
K-matrix form more than a qualitative difference in the
partial-wave amplitudes. We note that coupled-channel fits
have in the past~\cite{vrana} found the N(1535) width to be about
half the 200 MeV obtained in single-channel fits to eta
photoproduction~\cite{krusche}. 

The extracted $N(1520)$ $\eta N$ branching fraction is very small, as 
was expected.  Penner and Mosel~\cite{penner} found 0.0023$\pm$0.0004 
for this ratio using an older version of our $\pi N$ amplitudes as a 
representation of the $\pi N$ elastic scattering database.  A somewhat 
smaller value, 0.0008$\pm$0.0001 was found in a Mainz analysis of eta 
photoproduction data~\cite{tiator}.  We find this quantity to be 
rather sensitive to model details, but our range of values 
effectively spans the two previous determinations.  Fits~A and C have 
included the Crystal Ball data, and result in more precisely 
determined $\eta N$ branching fractions, as expected.  In fits B and 
D, the Crystal Ball data were excluded. Different $\rho N$ and 
$\pi\Delta$ branching fractions (Fits A and B versus C and D) were 
obtained through the choice of contributions to the background.  The 
background K-matrix contains elements coupling, for example, $\eta 
N\to\rho N$, which cannot be measured and are therefore intrinsically 
model-dependent.

\subsection{$\eta N$ scattering length}
\label{sec:slth}

\begin{figure*}[th]
\centering{
\includegraphics[height=0.45\textwidth, angle=90]{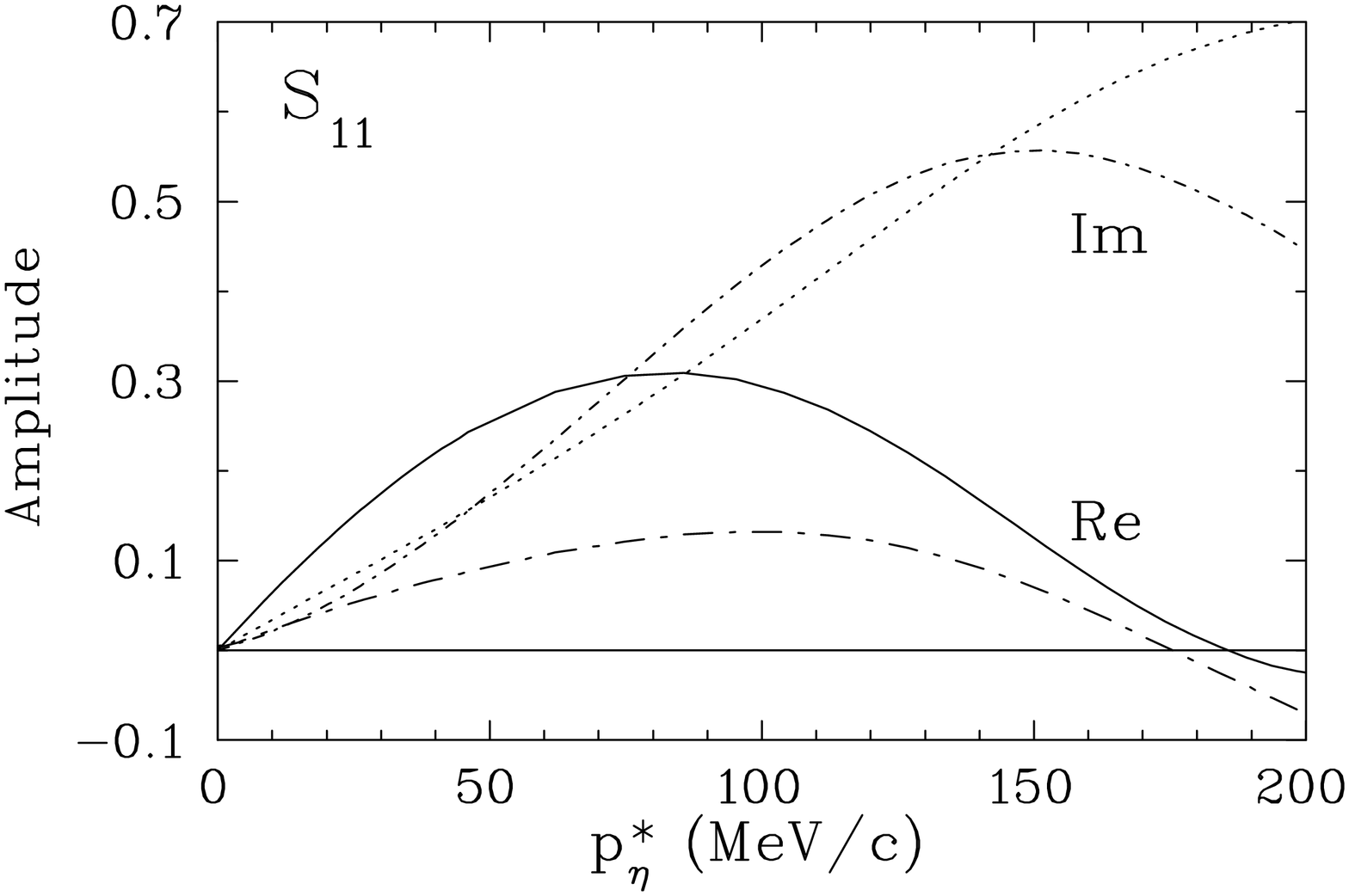}
}\caption{$p^{\ast}_{\eta}$ dependence of the $S_{11}$
          amplitude for the reaction $\eta n\to\eta n$.
          Dash-dotted (dotted) curves give the real (imaginary)
          parts of amplitudes corresponding to the solution G380.  
          Solid (short-dash-dotted) lines represent the 
          real (imaginary) parts of amplitudes Fit~A.  All 
          amplitudes are dimensionless.} \label{fig:a5}
\end{figure*}

As can be seen in Table~\ref{tbl1}, previous determinations of the 
$\eta N$ scattering length have produced widely varying results. 
This should not be surprising, as these determinations require the 
threshold behavior of an amplitude that cannot be directly measured. 
Somewhat surprising to us, however, was the relative stability of 
scattering lengths found from our set of four K-matrix fits, with 
and without the Crystal Ball data, which are shown in Table~\ref{tbl4}.  
These results are comparable to those found in a fit by Green and 
Wycech~\cite{green}, who used a similar K-matrix representation for 
the S-wave amplitudes, and the multi-channel fits of Penner and 
Mosel~\cite{penner}.   If, however, we determine the $\eta N$ 
scattering length directly from our global fit, based on a 
Chew-Mandlestam K-matrix formalism, a very different result is 
found. This value seems more compatible with the calculation of 
Ref.~\cite{kr01}. As a result, we can confirm previous 
determinations within similar approaches, but caution that (for 
the real part in particular) the employed model may be more 
important than improvements in the fitted data.  The S-wave $\eta n$ 
elastic partial waves, from G380 and Fit~A, are displayed in 
Fig.~\ref{fig:a5}.

\section{Conclusions and Future Prospects}
\label{sec:conc} 

We have explored the model and data dependence of $\eta N$
couplings to the $N(1535)$ and $N(1520)$ resonances, and have
extracted the $\eta N$ scattering length. Our values for these 
quantities are in reasonable agreement with previous 
determinations. One notable difference in our method has been
the direct fit to data, rather than to amplitudes. This has 
allowed a direct $\chi^2$ comparison of the fits. 

\begin{figure*}[th]
\centering{
\includegraphics[height=0.6\textwidth, angle=90]{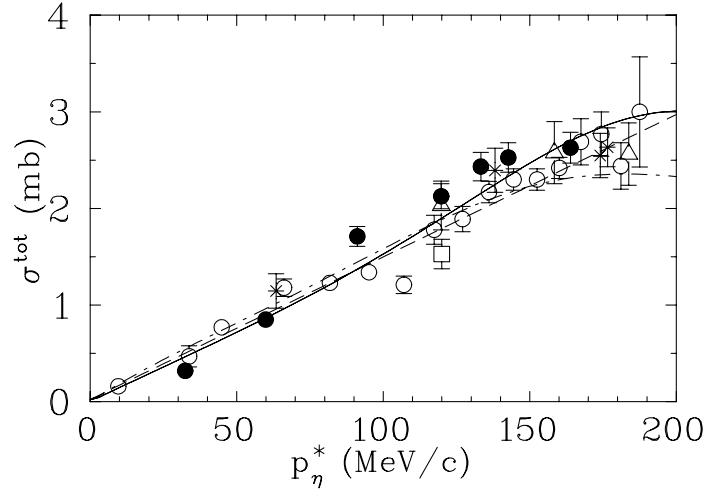}
}\caption{$p^{\ast}_{\eta}$ dependence of $\sigma^{tot}(\pi^-p
          \to\eta n)$.  Data and notation given in 
          Fig.~\protect\ref{fig:a1}.  Dashed line shows a linear 
          fit to E909~\protect\cite{e909} (open circles) data.}
          \label{fig:a2}
\end{figure*}

From an experimental point of view, several issues remain to be
resolved. The recent Crystal Ball measurements of $\pi^-p\to\eta 
n$, covering a region from threshold to the peak of the $N(1535)$ 
resonance, have suggested a slightly lower mass and width for this 
state. This could have been verified with measurements continuing 
to higher energies. However, with the Crystal Ball moved to Mainz, 
this is no longer possible. The standard value~\cite{pdg} 
(547.75$\pm$0.12~MeV) of the $\eta$ mass has also shifted recently, 
and this naturally effects extrapolations associated with the 
scattering length determination.  The linear plot in 
Fig.~\ref{fig:a2} has taken 547.3~MeV for the eta meson mass.  We 
have allowed the eta mass to vary between 547 and 548~MeV, finding 
very little sensitivity in our fits. We should also note that a 
recent measurement from the GEM collaboration~\cite{machner} finds 
a value close to the previous ``standard" mass of 547.3~MeV.

Given the model dependence found in our determinations, it would 
be interesting to see the effect of the Crystal Ball data in 
multi-channel fits which include representations of the $\pi 
N\to\pi\pi N$ data.  Also of interest would be a re-examination 
of the $N(1520)$ $\eta n$ branching fraction as extracted from 
eta-photoproduction data. The quality and quantity of data for 
this reaction has increased since the Mainz analysis~\cite{tiator}. 
Work on this subject is in progress.

\acknowledgments

This work was supported in part by the U.~S.~Department of Energy 
under Grants DE--FG02--95ER40901 and DE--FG02--99ER41110.
R.~A, I.~S., and R.~W. acknowledge partial support from Jefferson 
Lab, which is operated by the Southeastern Universities Research 
Association under DOE contract DE--AC05--84ER40150.  A.~G. 
acknowledges the hospitality extended by the Center for Nuclear 
Studies of The George Washington University.


\eject

\begin{table}[th]
\caption{$\eta N$-scattering length overview}
\label{tbl1}
\begin{center}
\begin{tabular}{|cc|cc|}
\hline
$A_{\eta N}$ (fm) & Reference & $A_{\eta N}$ (fm) & Reference \\
\hline
-0.15     +i~0.22       & \protect\cite{bi96} &  0.56      +i~0.22       & \protect\cite{bi96} \\
0.20      +i~0.26       & \protect\cite{ka97} &  0.577     +i~0.216      & \protect\cite{fe98} \\
      $\ge$i~0.24(2)    & \protect\cite{bi73} &  0.579     +i~0.399      & \protect\cite{kr94} \\
0.25      +i~0.16       & \protect\cite{be91} &  0.621(40) +i~0.306(34)  & \protect\cite{ab96} \\
0.27      +i~0.22       & \protect\cite{bh85} &  0.68      +i~0.24       & \protect\cite{ka95} \\
0.28      +i~0.19       & \protect\cite{bh85} &  0.717(30) +i~0.263(25)  & \protect\cite{ba97} \\
0.281     +i~0.360      & \protect\cite{kr94} &  0.734(26) +i~0.269(19)  & \protect\cite{ba98} \\
$\le$0.30               & \protect\cite{gr00} &  0.75(4)   +i~0.27(3)    & \protect\cite{gr97} \\
0.32      +i~0.25       & \protect\cite{ca00} &  $\ge$0.75               & \protect\cite{ra01} \\
0.404(117)+i~0.343(58)  & \protect\cite{ba95} &  0.75      +i~0.27       & \protect\cite{fi02} \\
0.42      +i~0.34       & \protect\cite{si02} &  0.772(5)  +i~0.217(3)   & \protect\cite{ni01} \\
0.42      +i~0.32       & \protect\cite{kr00} &  0.83      +i~0.35       & \protect\cite{tu65} \\
0.430     +i~0.394      & \protect\cite{kr94} &  0.87      +i~0.27       & \protect\cite{gr99} \\
0.46(9)   +i~0.18(3)    & \protect\cite{br02} &  0.876(47) +i~0.274(39)  & \protect\cite{ba95} \\
0.476     +i~0.279      & \protect\cite{fa95} &  0.886(47) +i~0.274(39)  & \protect\cite{ba95} \\
0.476     +i~0.279      & \protect\cite{ti94} &  0.91(6)   +i~0.27(2)    & \protect\cite{green} \\
0.487     +i~0.171      & \protect\cite{fe98} &  0.91(3)   +i~0.29(4)    & \protect\cite{ba95a} \\
0.51      +i~0.21       & \protect\cite{sa95} &  0.968     +i~0.281      & \protect\cite{ba95} \\
0.52      +i~0.25       & \protect\cite{wi97} &  0.980     +i~0.37       & \protect\cite{ar92} \\
0.54      +i~0.49       & \protect\cite{kr01} &  0.991     +i~0.347      & \protect\cite{penner} \\
0.55(20)  +i~0.30       & \protect\cite{wi93} &  1.05      +i~0.27       & \protect\cite{gr99} \\
0.550     +i~0.300      & \protect\cite{sa95} &                          & \\
\hline
\end{tabular}
\end{center}
\end{table}
\begin{table}[th]
\caption{The present (G380) and previous
         (FA02~\protect\cite{fa02} energy-dependent
         partial-wave analyses of elastic $\pi^\pm p$,
         charge-exchange ($\pi^0n$), and $\pi^-p\to\eta
         n$ ($\eta n$) scattering data, compared to fits
         A - D from 400 -- 900~MeV. 
         \label{tbl2}}
\begin{tabular}{|c|ccc|}
\colrule
Solution &
$\chi^2$/$\pi^-p$ & $\chi^2$/$\pi^0 n$ & $\chi^2$/$\eta n$  \\
\colrule
FA02  & 6286/2773 & 1920/1100 & 635/257 \\
G380  & 5825/2773 & 1723/1100 & 569/257 \\
Fit~A & 5961/2773 & 1684/1100 & 539/257 \\
Fit~B & 5935/2773 & 1748/1100 & 575/257 \\
Fit~C & 6001/2773 & 1732/1100 & 571/257 \\
Fit~D & 5961/2773 & 1839/1100 & 582/257 \\
\colrule
\end{tabular}
\end{table}
\begin{table}[th]
\caption{Resonance widths (in MeV) and branching fractions.
         \label{tbl3}}
\begin{tabular}{|c|c|ccccc|}
\colrule
Resonance &
Solution & $\Gamma_{\pi}$ & $\Gamma_{\eta}$ & $\Gamma_{\pi \Delta}$ &
$\Gamma_{\rho N}$ & $\Gamma_{\eta} / \Gamma_{\rm tot}$ \\
\colrule
\colrule
N(1535)& Fit~A & 30$\pm$2 & 45$\pm$3 & 15$\pm$1 &    & 0.50 \\
       & Fit~B & 32$\pm$3 & 45$\pm$4 & 16$\pm$1 &    & 0.48 \\
       & Fit~C & 39$\pm$3 & 67$\pm$4 &  9$\pm$2 &    & 0.58 \\
       & Fit~D & 42$\pm$6 & 70$\pm$10& 11$\pm$2 &    & 0.57 \\
\colrule
\colrule
N(1520)& Fit~A & 68$\pm$1 & 0.12$\pm$0.03 & 19$\pm$5 & 19$\pm$5 & 0.0012 \\
       & Fit~B & 68$\pm$1 & 0.17$\pm$0.12 & 19$\pm$6 & 19$\pm$6 & 0.0016 \\
       & Fit~C & 67$\pm$1 & 0.08$\pm$0.03 & 14$\pm$4 & 24$\pm$4 & 0.0008 \\
       & Fit~D & 67$\pm$1 & 0.09$\pm$0.07 & 14$\pm$5 & 24$\pm$5 & 0.0009 \\
\colrule
\end{tabular}
\end{table}
\begin{table}[t]
\caption{$\eta N$ scattering lengths from K-matrix fits 
         (resonance plus background, see text) and the global 
         energy-dependent fit (G380). 
         \label{tbl4}}
\begin{tabular}{|c|c|}
\colrule
Solution & Scattering Length~(fm) \\
\colrule
Fit~A & 1.14 + i 0.31 \\
Fit~B & 1.10 + i 0.30 \\
Fit~C & 1.12 + i 0.39 \\
Fit~D & 1.03 + i 0.41 \\
G380  & 0.41 + i 0.56 \\
\colrule
\end{tabular}
\end{table}


\begin{thebibliography}{99}
\bibitem{pdg} S.~Eidelman \textit{et al.} [Particle Data Group], 
              Phys.\ Lett.\ B\ \textbf{592}, 1 (2004).
\bibitem{penner} G.~Penner and U.~Mosel, Phys.\ Rev.\ C\ 
              \textbf{66}, 055211, (2002) [nucl-th/0207066].
\bibitem{tiator} L.~Tiator, D.~Drechsel, G.~Kn\"ochlein, and
              C.~Bennhold, Phys.\ Rev.\ C\ \textbf{60}, 035210 
              (1999) [nucl-th/9902028].
\bibitem{sokol}G.~A.~Sokol, A.~I.~L'vov, and L.~N.~Pavlyuchenko, 
              \textit{Proceedings of the Workshop on the Physics
              of Excited Nucleons (NSTAR2001), Mainz, Germany,
              March, 2001}, edited by D.~Drechsel and L.~Tiator 
              (World Scientific, 2001) p.~283 [nucl-ex/0106005].
\bibitem{bh85}R.~S.~Bhalerao and L.~C.~Liu, Phys.\ Rev.\ Lett.\ 
              \textbf{54}, 865 (1985);\\
              Q.~Haider and L.~C.~Liu, Phys.\ Lett.\ B\ 
              \textbf{172}, 257 (1986).
\bibitem{e909}T.~W.~Morrison \textit{et al.},
              Bull.\ Am.\ Phys.\ Soc.\ \textbf{45}, 58 (2000),\\
              T.~W.~Morrison, Ph.~D. Thesis, The George Washington 
              University, Dec.~1999.
\bibitem{pr05}S.~Prakhov \textit{et al.} [Crystal Ball      
              Collaboration], in press, Phys.\ Rev.\ C\ 
              \textbf{72}, (2005).
\bibitem{bi73}D.~M.~Binnie \textit{et al.}, Phys.\ Rev.\ D\ 
              \textbf{8}, 2789 (1973).
\bibitem{bu69}F.~Bulos \textit{et al.}, Phys.\ Rev.\ 
              \textbf{187}, 1827 (1969).
\bibitem{de75}N.~C.~Debenham \textit{et al.}, Phys.\ Rev.\ 
              D\ \textbf{12}, 2545 (1975).
\bibitem{de69}W.~Deinet \textit{et al.}, Nucl.\ Phys.\ 
              \textbf{B11}, 495 (1969).
\bibitem{ri70}B.~W.~Richards \textit{et al.}, Phys.\ Rev.\ 
              D\ \textbf{1}, 10 (1970).
\bibitem{kr95}B.~Krushe \textit{et al.}, Phys.\ Rev.\ Lett.\ 
              \textbf{74}, 3736 (1995).
\bibitem{pr95}J.~W.~Price \textit{et al.}, Phys.\ Rev.\ C\ 
              \textbf{51}, 2283 (1995).
\bibitem{re02}F.~Renard \textit{et al.} [GRAAL Collaboration], 
              Phys.\ Lett.\ B\ \textbf{528}, 215 (2002)
              [hep-ex/0011098].
\bibitem{dl69}B.~Delcourt \textit{et al.}, Phys.\ Lett.\ B\ 
              \textbf{29}, 75 (1969).
\bibitem{dy95}S.~A.~Dytman \textit{et al.}, Phys.\ Rev.\ C\ 
              \textbf{51}, 2710 (1995).
\bibitem{er68}R.~Erbe \textit{et al.}
              (Aachen-Berlin-Bonn-Hamburg-Heidelberg-M\"unchen 
              Collaboration), Phys.\ Rev.\ \textbf{175}, 1669 
              (1968).
\bibitem{du02}M.~Dugger \textit{et al.} [CLAS Collaboration], 
              Phys.\ Rev.\ Lett.\ \textbf{89}, 222002 (2002).
\bibitem{cr05}V.~Crede \textit{et al.} [CB--ELSA 
              Collaboration], Phys.\ Rev.\ Lett.\ \textbf{94}, 
              012004 (2005) [hep-ex/0311045].
\bibitem{bi96}B.~L.~Birbrair and A.~B.~Gridnev, Z.\ Phys.\ A\
              \textbf{354}, 95 (1996).
\bibitem{ka97}N.~Kaiser T.~Waas, and W.~Weise, Nucl.\ Phys.\ 
              \textbf{A612}, 297 (1997) [hep-ph/9607459].
\bibitem{be91}C.~Bennhold and H.~Tanabe, Nucl.\ Phys.\ 
              \textbf{A530}, 625 (1991).
\bibitem{kr94}B.~Krusche, in \textit{Proc. of the II TAPS 
              Workshop, Guadamar, Spain, 1993}, edited by J.~Diaz 
              and Y.~Schutz (World Scientific, 1994), p.~310.
\bibitem{gr00}V.~Yu.~Grishina \textit{et al.}, Phys.\ Lett.\ B\ 
              \textbf{475}, 9 (2000) [nucl-th/990549].
\bibitem{ca00}J.~Caro~Ramon \textit{et al.}, Nucl.\ Phys.\ 
              \textbf{A672}, 249 (2000) [nucl-th/9912053].
\bibitem{ba95}M.~Batinic \textit{et al.}, Phys.\ Rev.\ C\ 
              \textbf{52}, 2188 (1995) [nucl-th/9502017].
\bibitem{si02}A.~Sibirtsev \textit{et al.}, Phys.\ Rev.\ C\ 
              \textbf{65}, 044007 (2002) [nucl-th/0111086].
\bibitem{kr00}O.~Krehl \textit{et al.}, Phys.\ Rev.\ C\ 
              \textbf{62}, 025207 (2000) [nucl-th/9911080].
\bibitem{br02}W.~Briscoe \textit{et al.}, \textit{Proceedings 
              of the 9th International Symposium on Meson-Nucleon 
              Physics and the Structure of the Nucleon (MENU2001), 
              Washington, DC, USA, July, 2001}, edited by 
              H.~Haberzettl and W.~J.~Briscoe, $\pi N$Newslett, 
              \textbf{16}, 391 (2002).
\bibitem{fa95}G.~F\"aldt and C.~Wilkin, Nucl.\ Phys.\ \textbf{A587}, 
              769 (1995).
\bibitem{ti94}L.~Tiator \textit{et al.}, Nucl.\ Phys.\ 
              \textbf{A580}, 455 (1994) [nucl-th/9404013].
\bibitem{fe98}T.~Feuster and U.~Mosel, Phys.\ Rev.\ C\ \textbf{58}, 
              457 (1998) [nucl-th/9708051].
\bibitem{sa95}Ch.~Sauermann, B.~L.~Frieman, and W.~N\"orenberg, 
              Phys.\ Lett.\ B\ \textbf{341}, 261 (1995) 
              [nucl-th/9408012];\\
              Ch.~Deutsch-Sauermann, B.~L.~Frieman, and 
              W.~N\"orenberg, Phys.\ Lett.\ B\ \textbf{409}, 51 
              (1997) [nucl-th/9701022].
\bibitem{wi97}N.~Willis \textit{et al.}, Phys.\ Lett.\ B\ 
              \textbf{406}, 14 (1997) [nucl-ex/9703002].
\bibitem{kr01}B.~Krippa, Phys.\ Rev.\ C\ \textbf{64}, 047602 
              (2001).
\bibitem{wi93}C.~Wilkin, Phys.\ Rev.\ C\ \textbf{47}, R938 
              (1993) [nucl-th/9301006].
\bibitem{ab96}V.~V.~Abaev and B.~M.~K.~Nefkens, Phys.\ Rev.\ C\ 
              \textbf{53}, 375 (1996).
\bibitem{ka95}N.~Kaiser, P.~B.~Siegel, and W.~Weise, Phys.\ Lett.\
              B\ \textbf{362}, 23 (1995) [nucl-th/9507036];\\
              N.~Kaiser \textit{et al.}, Nucl.\ Phys.\ 
              \textbf{A594}, 325 (1995) [nucl-th/9505043].
\bibitem{ba97}M.~Batinic \textit{et al.}, nucl-th/9703023.
\bibitem{ba98}M.~Batinic \textit{et al.}, Phys.\ Scr.\ \textbf{58}, 
              15 (1998).
\bibitem{gr97}A.~M.~Green and S.~Wycech, Phys.\ Rev.\ C\ \textbf{55}, 
              R2167 (1997) [nucl-th/9703009].
\bibitem{ra01}S.~A.~Rakityansky \textit{et al.}, Nucl.\ Phys.\ 
              \textbf{A684}, 383 (2001);\\
              N.~V.~Shevchenko \textit{et al.}, Phys.\ Rev.\ C\ 
              \textbf{58}, 3055 (1999) [nucl-th/9808009].
\bibitem{fi02}A.~Fix and H.~Arenh\"ovel, Phys.\ Rev.\ C\
              \textbf{66}, 024002 (2002).
\bibitem{ni01}J.~Nieves and E.~Ruiz~Arriola, Phys.\ Rev.\ D\ 
              \textbf{64}, 116008 (2001) [hep-ph/0104307].
\bibitem{tu65}S.~F.~Tuan, Phys.\ Rev.\ \textbf{139B}, 1393 (1965).
\bibitem{gr99}A.~M.~Green and S.~Wycech, Phys.\ Rev.\ C\ 
              \textbf{60}, 035208 (1999) [nucl-th/9905011].
\bibitem{green}A.~M.~Green and S.~Wycech, Phys.\ Rev.\ C\ 
              \textbf{71}, 014001 (2005) [nucl-th/0411024].
\bibitem{ba95a}M.~Batinic \textit{et al.}, Phys.\ Rev.\ C\
              \textbf{51}, 2310 (1995) [nucl-th/9501011].
\bibitem{ar92}M.~Arima \textit{et al.}, Nucl.\ Phys.\ 
              \textbf{A543}, 613 (1992).
\bibitem{kh80}G.~H\"ohler, \textit{Pion--Nucleon Scattering},
              Landoldt--B\"ornstein Vol. \textbf{I/9b2}, edited
              by H.~Schopper (Springer Verlag, 1983).
\bibitem{sm95}R.~A.~Arndt, R.~L.~Workman, I.~I.~Strakovsky,
              and M.~M.~Pavan, Phys.\ Rev.\ C \textbf{52},
              2120 (1995) [nucl-th/9505040].
\bibitem{fa02}R.~A.~Arndt, W.~J.~Briscoe, I.~I.~Strakovsky,
              R.~L.~Workman, and M.~M.~Pavan, Phys.\ Rev.\ C\
              \textbf{69}, 035213 (2004) [nucl-th/0311089].
\bibitem{isaid}The full database and numerous PWAs can be
               accessed at the website 
               \hbox{http://gwdac.phys.gwu.edu}.
\bibitem{clnef}M.~Clajus and B.~M.~K.~Nefkens, $\pi N$~Newslett.
               \textbf{7}, 76 (1992).
\bibitem{vrana} T.~P.~Vrana, S.~A.~Dytman, and T.~S.~H.~Lee,
                Phys.\ Rept.\ \textbf{328}, 181 (2000)
                [nucl-th/9910012].
\bibitem{krusche}B.~Krusche, N.~C.~Mukhopadhyay, J.~F.~Zhang, and
                M.~Benmerrouche, Phys.\ Lett.\ B\ \textbf{397}, 
                171 (1997).
\bibitem{machner}M. Abdel-Bary \textit{et al.} [GEM Collaboration],
                ``A precision determination of the mass of the $\eta$
                meson", [hep-ex/0505006].
\end{thebibliography}
\end{document}